\documentclass[12pt,preprint]{aastex}
\usepackage{graphicx}
\usepackage{amsmath}
\usepackage{natbib}
\newcommand\clock{\count0=\time \divide\count0 by 60
     \count1=\count0 \multiply\count1 by -60 \advance\count1 by \time
     \number\count0:\ifnum\count1<10{0\number\count1}\else\number\count1\fi}

\shortauthors{Christensen et al.}
\shorttitle{SF and Feedback in SPH Simulations II}

\begin{document}

\title{Star Formation and Feedback in Smoothed Particle Hydrodynamic Simulations II: Resolution Effects}

\author{Charlotte R. Christensen 
and Thomas Quinn
}
\affil{Dept. of Astronomy, Univ. of Washington, Box 351580, Seattle WA, 98195}
\email{christensen@astro.washington.edu}

\author{Gregory Stinson
}
\affil{Jeremiah Horrocks Institute, University of Central Lancashire, Preston, PR1 2HE UK}

\author{Jillian Bellovary
}
\affil{Dept. of Astronomy, Univ. of Washington, Box 351580, Seattle WA, 98195}

\author{James Wadsley
}
\affil{Dept. of Physics \& Astronomy, ABB-241, McMaster Univ., 1280 Main St. W, Hamilton, ON, L8S 4M1, Canada}

\newpage
  
\begin{abstract}
We examine the effect of mass and force resolution on a specific star formation (SF) recipe using a set of \emph{N}-body/Smooth Particle Hydrodynamic simulations of isolated galaxies.
Our simulations span halo masses from $10^9$ to $10^{13}$ $M_{\odot}$, more than four orders of magnitude in mass resolution, and two orders of magnitude in the gravitational softening length, $\epsilon$, representing the force resolution.
We examine the total global star formation rate, the star formation history, and the quantity of stellar feedback and compare the disk structure of the galaxies.
Based on our analysis, we recommend using at least $10^4$ particles each for the dark matter and gas component and a force resolution of $\epsilon \approx 10^{-3} R_{vir}$ when studying global SF and feedback.
When the spatial distribution of stars is important,  the number of gas and dark matter particles must be increased to at least $10^5$ of each.
Low mass resolution simulations with fixed softening lengths show particularly weak stellar disks due to two-body heating.
While decreasing spatial resolution in low mass resolution simulations limits two-body effects, density and potential gradients cannot be sustained.
Regardless of the softening, low-mass resolution simulations contain fewer high density regions where SF may occur.
Galaxies of approximately  $10^{10}$ $M_{\odot}$ display unique sensitivity to both mass and force resolution.  
This mass of galaxy has a shallow potential and is on the verge of forming a disk.
The combination of these factors give this galaxy the potential for strong gas outflows driven by supernova feedback and make it particularly sensitive to any changes to the simulation parameters. 
\end{abstract}

\section{Introduction}  \label{introsec}

The formation of stars is a highly non-linear process that is dependent on a number of different physical mechanisms, possibly including but not limited to: gravitational \citep{Zinnecker84, Larson85} or turbulent \citep{PadoanANDNordlund02, BateANDBonnell04} fragmentation, competitive accretion \citep{BonnellANDBate06}, magnetic fields \citep{Shu04}, stellar feedback \citep{Silk95, AdamsANDFatuzzo96}, and the coalescence of protostars \citep{MurrayANDLin96}.
On larger scales, SF is driven by gravitational interactions between galaxies \citep{Barton00, Li08}, external gas accretion \citep{Kauffmann06}, and disk instabilities such as bars \citep{Ho97, Sheth05} and spiral arms \citep{Roberts69, Tamburro08}.
Furthermore, 
energy from active galactic nuclei and supernova feedback \citep{Martin99, Kaviraj07, Bundy08}
regulate SF.
Hence, SF is determined by the interaction between many large and small scale physical processes.

SF is commonly studied on a galactic scale using \emph{N}-body/Smoothed Particle Hydrodynamic (SPH) simulations because of their ability to model processes over orders of magnitude in space and mass while automatically resolving denser areas to a higher degree.
In fact, it is only very recently that an Eulerian hydrodynamic simulation has followed the formation of a galaxy in a cosmological context to a redshift of zero \citep{Gibson09}.
Cosmological SPH simulations \citep{Keres05, Governato07, Brooks09} contain one or more galaxies forming within a larger, cosmologically determined dark matter (DM) structure and are crucial for replicating the tidal interactions with other galaxies that may trigger SF \citep{Okamoto05, Kaufmann06, Governato07}.
These cosmological simulations are used for studying the connection between SF and UV background radiation \citep{Bullock00, Kravtsov04, Moore06}, mergers \citep{Abadi03a}, tidal and ram pressure stripping \citep{Mayer06}, and gas accretion by galaxies \citep{Keres05, Brooks09}.  
On a smaller scale, simulations of isolated galaxies give a resolved model of SF within a galaxy and its effect on the galactic structure \citep{Valenzuela03, Debattista04, Tasker06, Stinson07, Roskar08}.

Even in high resolution SPH simulations of the smallest isolated galaxies, the formation of individual stars is unresolved.
In order to describe the non-linear, sub-resolution process in SPH simulations, SF is implemented using sub-grid models wherein the complex physical processes are averaged over the scale of the smallest resolution element.
Comparisons with Eulerian grid models of SF in the interstellar medium demonstrate the viability of sub-grid SF methods in SPH simulations \citep{Tasker06}. 
Given basic information about the surrounding gas, such as density and temperature, statistical models of SF can reproduce most observed global SF properties of galaxies in well-resolved simulations.
The question remains, however, what is necessary for a simulation to be ``well-resolved"?
Although the qualitative behavior of SF in SPH simulations has been discussed in previous papers, here we will establish the minimum spatial and mass resolution necessary to model different aspects of SF in galaxies simulated with the widely used code GASOLINE \citep{Wadsley04} over a wide range of masses.
As well as delineating the behavior of SF with GASOLINE as a function of resolution, these results  may be applicable to other SPH simulations in which individual gas particles represent single-phases of the ISM, SF occurs in suitably cold and dense gas particles, and energy from supernova is transferred to nearby gas particles, e.g. \citet{Katz96, ThackerCouchman01, Keres05, Saitoh08}.

 \citet{Stinson06}, hereafter S06, described a specific sub-grid model for SF and a blastwave model for stellar feedback in GASOLINE, laid out a physical justification for the models, and presented several different tests of the SF and feedback models.
To judge the effects of different SF criteria and model parameters, S06 used an isolated Milky Way-mass disk galaxy.
When using the best-fit parameters, this model galaxy was able to reproduce the Schmidt law, $\Sigma_{SFR}\propto \Sigma_{gas}^{1.4}$ \citep{Kennicutt98}, with a near constant star formation rate (SFR).  
S06 further simulated this isolated galaxy at three different mass resolutions ($2000$, $10 000$ and $50 000$ initial gas particles and a gas disk mass of $10^9  M_{\odot}$). 
While fewer stars formed in simulations run at lower resolution, the global SFR started to converge for $10 000$ initial gas particles.
\citet{Brooks07} found that the total amount of SF and the metallicity converged for over a few thousand DM particles per galaxy in cosmological simulations.
\citet{Governato07} further tested this SF recipe when calibrating the SF parameters through comparisons of the SFR and velocity dispersion of an isolated dwarf and a MW-mass galaxy to observations.
This SF and feedback recipe, albeit using a different set of parameters, has been successful in reproducing the Tully-Fisher Relation \citep{Governato07},  the internal structure of disk galaxies \citep{Governato04, Governato07, Roskar08, Stinson09}, the column density distributions of damped Layman $\alpha$ systems \citep{Pontzen08,Pontzen10}, and the observed stellar mass-metallicity relationship \citep{Brooks07}.

Despite this success, it is possible that SF in galaxy simulations outside of the previously tested range of resolution could behave inconsistently.
\citet{Saitoh08} found that when using a similar SPH/N-body code, in very high resolution simulations ($3.5 \times 10^2 - 10^3 M_\odot$ per particle), increasing the threshold density for SF and decreasing the efficiency resulted in more realistic disks which were both thinner and clumpier.
If simulations with different resolutions show different properties, such as a range of gas densities, the SF recipe may need to be adjusted at these extremes.
As well as affecting the robustness of the particular sub-grid SF model used, resolution also affects the dynamics of galaxy simulations and impacts the SF through the alteration of the density structure of the galaxy \citep{SpringelANDHernquist02, Governato07, Kaufmann07, Naab07, Foyle08}.
Unlike the recipe put forth in \citet{SpringelANDHernquist03a}, which includes a sub-resolution multiphase model of the ISM, the S06 recipe assumes a single pressure, density and temperature for the entire gas particle.
\citet{SpringelANDHernquist03a} developed their multiphase model in part to minimize resolution effects but at the cost of small scale spatial structure of the multi-phase ISM.
Because GASOLINE uses a single phase ISM for each particle it is especially important that the density structure of the ISM be well resolved to correctly model the gas associated with SF.
Understanding the impact of resolution becomes particularly important when analyzing large-volume, cosmological simulations that contain a number of galaxies with different masses, each resolved to different degrees.
The wide range of galactic masses in these cosmological simulations mean SF must remain accurate in a variety of different galactic environments, from large galaxies with high densities of gas to small ones from which gas is easily expelled.
It is, therefore, necessary to establish how SF behaves at different mass and spatial resolutions over a range of galactic masses.

There is a large body of work on the effect of both mass resolution (the mass of the particles and inversely related to the number of particles) and force resolution (the gravitational softening length, $\epsilon$) on the dynamics of SPH simulated galaxies.
The effects of insufficient resolution can be largely grouped as those relating to over-cooling \citep{HutchingsANDThomas00,SpringelANDHernquist02}, two-body heating \citep{Steinmetz97, Governato07,Mayer08}, artificial viscosity \citep{Kaufmann07, Wadsley08}, and small-scale gravitational support \citep{BateANDBurkert97}.
The first two effects primarily relate to the mass resolution whereas the third is connected to the force resolution.
These effects can make the galaxy either more or less conducive to SF and they can even impact each other.
Understanding how each alone impacts the dynamics of a simulated galaxy, however, is the first step to understanding how together they may influence SF.

The first potential problem for accurately modeling SF in low resolution SPH simulations is that over-cooling of gas in galactic halos simulated with small numbers of particles \citep{SpringelANDHernquist02} may allow the gas to reach the density threshold for SF too quickly.  
This cooling, also known as in-shock cooling \citep{HutchingsANDThomas00} happens when low resolution causes the broadening of shock fronts and contact discontinuities and, as a result, the gas moving through them is heated relatively slowly.  
This longer heating time-scale gives the gas the opportunity to cool radiatively, which reduces the post-shock cooling time by as much as a factor of two and lowers the maximum shock temperature by up to $50\%$ in shock-tube simulations \citep{HutchingsANDThomas00}.

In low mass resolution simulations, two-body heating and the processes related to it may affect SF by changing both the dynamics and the gas physics.
Two-body heating results from approximating a smooth potential with discrete masses and causes an artificial increase of kinetic energy. 
In particular, the presence of  large collisionless bodies, such as dark matter (DM) particles, can heat the disk and destroy galactic disk structure \citep{LaceyANDOstriker85}.
Even when the rotational disk is not fully degraded through two-body heating, numerical relaxation can transfer angular momentum from the disk to the halo \citep{Kaufmann07}.
This results in a weaker disk component \citep{Governato04} and excessively concentrated disks \citep{Governato07}.
In addition to changing the dynamics, two-body heating increases the gas temperature through artificial viscous dissipation \citep{Steinmetz97}, making it increasingly difficult for gas to reach temperatures below the temperature threshold for SF.

One final way in which the small number of particles in a simulation may affect SF is through 
the reduced spatial resolution of the gas. 
Simulations with small numbers of particles have their gas properties calculated over larger smoothing lengths, $h$, which in turn adds artificial viscosity and dissipation \citep{Kaufmann07} and results in an increased loss of angular momentum and smaller, more concentrated disks.
The smoothing of gas properties over larger volumes also results in fewer regions with sufficiently cool temperatures and high densities for SF.

The ability of regions to collapse to form stars in \emph{N}-body simulations is likewise sensitive to the force resolution, i.e. the gravitational softening length.
Large values of $\epsilon$ spread the potential over a larger volume, resulting in decreased density fluctuations and producing puffier disks \citep{Kaufmann07}.
Increasing $\epsilon$, therefore, decreases the amount of gas able to reach sufficient densities for SF.
Larger values of $\epsilon$ also hinder the formation of bars as bars are only formed in the non-axisymmetric modes of the density field that are larger than $\epsilon$.
Without bars, spiral structure is damped and less angular momentum is transferred outward \citep{Kaufmann07} causing a lower concentration. 
Without the high densities of gas produced in bars, spiral arms, and in the center of the galaxy, SF may be reduced.

When the Jeans length is comparable to either $\epsilon$ or $h$ the relationship between them becomes particularly important.
When $\epsilon \approx h$, gravitational forces on scales smaller than the softening length are reduced, causing artificial support.
Conversely, when $\epsilon < h$ gravitational forces dominate over pressure forces \citep{BateANDBurkert97}.
Reducing $\epsilon$ relative to $h$ increases the relative gravitational importance, while increasing it results in greater pressure support.
Once the Jeans length begins to be unresolved, gas dynamics, and through that SF, become dependent on the relative values of $\epsilon$ and $h$  \citep{BateANDBurkert97}.
The extent to which relative values of $\epsilon$ and $h$ affect SF is itself dependent on the implementation of the SF recipe. 
For this reason, it is important that the Jeans mass and length be greater than the respective mass and softening lengths of the star forming particles.

As well as the effects of resolution on SF, the effects on the supernova feedback that regulates it must also be understood.
All SPH feedback recipes must contend with the feedback energy transferred to the nearby dense gas being radiated away too quickly, resulting in little hydrodynamical effect \citep{Katz96}.
One commonly used solution is to inject some of the SN energy in the form of kinetic rather than thermal energy  to ``wind particles'' \citep{SpringelANDHernquist03a}.
Energy from feedback not used in the wind is necessary to maintain the equation of state in the presence of radiative losses.
The effect of mass resolution on the wind produced in this recipe have been examined by \citet{DallaVecchia08}, among others.
A second solution to simulating the sub-grid adiabatic phase of supernova explosions is to temporarily disabling gas cooling.
This is the approach taken by the blastwave model, first presented in \citet{Gerritsen97} and \citet{ThackerCouchman00} and later developed for Gasoline in S06. 
As no kinetic energy is explicitly transferred in this recipe, resolution will have a different effect.
For example, in low mass resolution simulations using the S06 recipe, supernova events occur less frequently but are more powerful.
In such simulations, the energy released per event is greater, that energy is distributed over fewer particles, and the cooling for each particle is disabled for a greater period of time (S06).
While this recipe produces the same galaxy-and-time-averaged quantities irrespective of resolution, (for instance, cooling is on average disabled for the same total mass of particles in a high resolution simulation as in a lower resolution one at any given time), it is an open question as to whether the different spatial and temporal distributions of supernova events in low and high mass resolution simulations will produce structurally different galaxies.
Furthermore, the high spatial detail produced by using large numbers of particles may allow for the formation of non-spherically symmetric structures, such as galactic winds.
Such structures would likewise be affected by resolution induced changes to the artificial viscosity of gas particles \citep{Kaufmann07} and changes to the disk scale height and the geometry of the disk \citep{MacLow88, MacLow89}.

In this paper, we explore the effects of force and mass resolution on S06 SF and stellar feedback in different mass galaxies.
When analyzing galaxies with this SF recipe, it is necessary to know the lower limit of mass and force resolution above which SF and the stellar disk properties converge.
Correctly determining the amount and location of SF is vital in simulations in galaxies because SF 1) produces and distributes metals throughout the galaxy, 2) affects the distribution of matter in the galaxy through feedback, and 3) enables us to relate simulations to observations.
Therefore, we ask: what resolutions are sufficient to establish the galactic SF rates, the history of SF, the amount of stellar feedback, and the distribution of stellar and gaseous matter?
To this end, we simulated a series of isolated galaxies of different masses at different mass and force resolutions and analyzed properties relating to SF.
Isolated galaxies have the advantage of requiring less computational expense while allowing us to separate out environmental effects from those caused by different resolutions.
We then analyze the SF and stellar feedback in the models and relate them to the history and structure of the galaxies to determine the resolution necessary for convergence and to describe the effects of resolution.

The outline of our paper is as follows.
In \S 2, we introduce the models and describe the computational methods used.
The results of these simulations are described in terms of global SF (\S 3.1), stellar feedback (\S 3.2) and stellar distribution (\S 3.3).
We address the physical connection between the SF, feedback, and structure and discuss implications of these results for large volume simulations in \S 4.  We conclude \S 5 with a list of recommendations for future cosmological simulations.

\section{Methods and Models}\label{methodssec}
Our set of isolated galaxies ranges in mass from $10^9 M_\odot$ 
to $10^{13} M_\odot$.
We simulated each galaxy at five mass resolutions with a range of 50 to $10^5$ DM particles and with the same number of initial gas particles.
For simplicity, we refer to the mass resolution of a given simulation by the number of DM particles with the understanding that this also denotes the number of gas particles initially in the simulation.
Based on our initial conditions, the mass of individual dark matter particles is $0.9 \times M_{H}/N_{DM}$ whereas the mass of each gas particle is $0.1 \times M_{H}/N_{DM}$, where $M_{H}$ is the total halo mass and $N_{DM}$ is the number of dark matter particles.
We chose to further focus on the $10^{12} M_\odot$ and the $10^{10} M_\odot$ galaxies.  
The former is representative of a Milky Way-mass galaxy.  
The latter is on the cusp of being able to form a disk through purely radiative cooling of primordial gas and, as will be shown, is particularly sensitive to SF and stellar feedback.
In addition to the 50 to $10^{5}$ DM particle runs, we simulated these two galaxies using $10^{6}$ DM particles to check convergence at very high mass resolution.
To examine the effect of force resolution, we re-simulated the $10^{12} M_\odot$ and the $10^{10} M_\odot$ simulations with $10^{5}$ DM particles and a range of softening lengths between $2.0\times 10^{-2}$  and $2.5\times 10^{-4} R_{vir}$, where $R_{vir}$ is the viral radius, defined by the radius where the enclosed density is 200 times the critical density.
A complete list of the simulations can be found in Table 1.

\begin{deluxetable}{lcr}
\tablecolumns{3}
\tablewidth{0pc}
\tablecaption{Simulations used in our analysis}
\tablehead{
\colhead{Mass [$M_\odot$]} & \colhead{Softening Length [$R_{vir}$]} & \colhead{Softening Length [pc]}  
}
\startdata
$10^{9} $                  & $1.0 \times10^{-3}$                                 &   21   \hspace{0.75cm}    \\ 
$10^{10}$*                 & $1.0\times 10^{-3}$                                 &   44   \hspace{0.75cm}    \\
\hspace{0.75cm} $10^{10}$                     &  \hspace{1.5cm} $2.5\times 10^{-4}$                                 &   11  \\
\hspace{0.75cm} $10^{10}$                     & \hspace{1.5cm} $2.5\times 10^{-3}$                                  &  110  \\
\hspace{0.75cm} $10^{10}$                     & \hspace{1.5cm} $5.0\times 10^{-3}$                                  &  220  \\
\hspace{0.75cm} $10^{10}$                     & \hspace{1.5cm} $1.0\times 10^{-2}$                                  &  440  \\
\hspace{0.75cm} $10^{10}$                     & \hspace{1.5cm} $2.0\times 10^{-2}$                                  & 1760  \\ 
$10^{11} $                 & $1.0\times 10^{-3}$                                 &   96   \hspace{0.75cm}   \\ 
$10^{12} $*                & $1.0\times 10^{-3}$                                 &  206   \hspace{0.75cm}   \\
\hspace{0.75cm} $10^{12} $                     & \hspace{1.5cm} $2.5\times 10^{-4}$                                 & 51.5  \\
\hspace{0.75cm} $10^{12} $                     & \hspace{1.5cm} $2.5\times 10^{-3}$                                 &  515  \\
\hspace{0.75cm} $10^{12} $                     & \hspace{1.5cm} $5.0\times 10^{-3}$                                 & 1030  \\
\hspace{0.75cm} $10^{12} $                     & \hspace{1.5cm} $1.0\times 10^{-2}$                                 & 2060  \\
\hspace{0.75cm} $10^{12} $                     & \hspace{1.5cm} $2.0\times 10^{-2}$                                 & 8240  \\  
$10^{13} $                  & $1.0\times 10^{-3}$                               &   444   \hspace{0.75cm}   \\  \hline
\enddata

\tablecomments{We simulated each of the listed galaxies with $50$, $100$, $10^3$, $10^4$ and $10^5$ DM particles and an equal number of initial gas particles.  
Galaxies marked with an asterisk were additionally simulated with $10^6$ DM particles and an equal number of initial gas particles.
For both the $10^{12} M_\odot$ and the $10^{10} M_\odot$ galaxies, we additionally ran a suite of simulations with various softening lengths.  The simulations with softening lengths not equal to $1.0 \times 10^{-3} R_{vir}$ are distinguished by the indented columns.}
\end{deluxetable}

\subsection{Isolated Galaxy Model}
The initial conditions for our isolated galaxies are identical to those used in \citet{Kaufmann07} and \citet{Stinson07}. 
They consist of an equilibrium, live DM halo \citep{Kazantzidis04}, with a concentration of $c = 8$, where $c = R_{vir}/R_{s}$ and $R_{s}$ is the scale radius of the NFW profile.
In order to establish equilibrium, the DM halo extends beyond $R_{vir}$ at a slightly lower mass resolution, such that 72\% of the DM mass and 90\% of the DM particles are within the virial radius.  
The DM velocity distribution was set using the Eddington inversion method of Kazantzidis et al. 2004.
Once the DM halo was created, $10\%$ of the mass of the DM halo was converted to gas and distributed evenly over the same number of gas particles as DM particles.
The gas was given the same NFW density profile as the DM and the temperature of each particle was chosen to establish hydrostatic equilibrium prior to cooling.
The gas cloud was then spun with a uniform rotational velocity such that $\lambda \equiv (j_{gas}|E^{1/2}|)(G M^{3/2}) = 0.039$, where $j_{gas}$ is the average specific angular velocity of the gas and $E$ and $M$ are the total energy and mass of the halo. 
By creating our isolated galaxies from collapsing clouds of gas such as these, we avoid requiring the galaxies to be in equilibrium and enable the study of both a sudden surge of SF during the initial collapse and sustained SF later on.

\subsection{Numerics}
Our simulations were performed using GASOLINE, a parallel tree+SPH code \citep{Wadsley04}, which is an extension of the \emph{N}-body gravity code PKDGRAV \citep{Stadel01}.
A softening length, $\epsilon$, is included in the calculation of gravitational force to decrease relaxation from close encounters of particles and to make the simulation numerically feasible.
We use spline kernel gravitational softening \citep{HernquistANDKatz89}.
The force on each particle is integrated using a timestep  $\Delta t_{grav} = \eta \sqrt{\frac{\epsilon}{a}}$, where the timestep criterion $\eta = 0.175$ and $a$ is the acceleration.
For gas particles, the timestep also satisfies the Courant condition \citep{Wadsley04}. 
Each gas particle has an SPH smoothing radius, $h$, over which gas properties are sampled.
The value for $h$ is set for each gas particle such that above a minimum value of $h \propto \epsilon$, the same number of particles are contained in each smoothing kernel; when $h$ is equal to the minimum value, more particles may be contained within the kernel.
In our simulations $h \geq 0.1 \epsilon$ and 32 particles are contained in each kernel when $h > 0.1\epsilon$.

We implement gas cooling assuming ionization equilibrium and an ideal gas of primordial composition.
The cooling rate and the ion abundances are solved for using collisional ionization rates \citep{Abel97}, radiative recombination \citep{Black81, VernerANDFerland96}, bremsstrahlung, and line cooling \citep{Cen92}.
As in S06, the effects of the UV background radiation, metal line cooling and molecular hydrogen cooling are not included. 

We modeled SF using the sub-grid recipe outlined in S06.  
The formation of each star particle (representing a simple stellar population of a set metallicity) is a statistical process based on the properties of the local gas.
In this recipe, gas particles must meet the following set of criteria to be capable of forming stars: the gas density must be greater than a minimum density, $n_{min}$, the temperature must be less than a maximum temperature, $T < T_{max} = 15,000$K, 
and the gas must be in a converging flow \citep{Katz92}.
Whether a potentially star forming gas particle spawns a star particle in a time step, $\Delta t$, is determined stochastically with the following probability $p$:
\begin{equation}
p = \frac{m_{gas}}{m_{star}}(1 - e^{-c^* \Delta t /t_{form}})
\end{equation}
in which $m_{gas}$ is the mass of the gas particle, $m_{star}$ is the initial mass of the potential star particle, $c^*$ is a constant star forming efficiency factor (the likelihood a valid gas particle will make a star particle in a formation time), and the formation time, $t_{form}$, is equal to the dynamical time (S06).
We chose $m_{star}$ to be 40\% of the initial gas particle mass as a compromise between the resolution of the galaxy's stellar component and computational expense.
This set of minimum star forming criteria and stochastic SF parameters leaves two tunable parameters: $n_{min}$ and $c^*$. 
The tests in \citet{Governato07} determined best values of $n_{min} = 0.1 \mathrm{cm}^{-3}$ and $c^* = 0.05$ and we use a \citet{Kroupa93} initial mass function to our star particles, as they did. 

Feedback from star particles represents the energy injected into the interstellar medium (ISM) by Type Ia and Type II supernovae.  
To calculate the effect of the released energy, we use the blastwave recipe described in S06 based on work of \citet{McKee77}.
Energy from a given star particle is equal to the number of supernovae that would occur in a simple stellar population of the same mass and metallicity multiplied by a constant value, $E_{SN} = 4\times 10^{50}$ ergs per supernova (S06). 
The energy is then distributed through the surrounding gas particles and the cooling is disabled for a length of time and distance determined by the blastwave model.
Feedback from stellar particles also returns mass and metals to the ISM.

\section{Results}  \label{datasec}
\subsection{Global Star Formation}
The global SFR of a galaxy (the mass of stellar particles formed in one year in the entire galaxy) is important to establish in part because it can be ascertained even in relatively low-resolution observational surveys of galaxies \citep{Bell03, Brinchmann04} and from observations of high-redshift galaxies \citep{Madau98, Papovich01, Shapley01, Stanway03}.  Furthermore, the global SFR is key to the integrity of the simulation because of the significant impact of stellar feedback on the structure of the galaxy.
Through the injection of metals and energy by stellar feedback, gas content, chemical evolution, morphology, and future SFR of the galaxy are all dependent on the global SFR.

\begin{figure}[ht]
\begin{center}$
\begin{array}{cc}
\includegraphics[width = .45\textwidth]{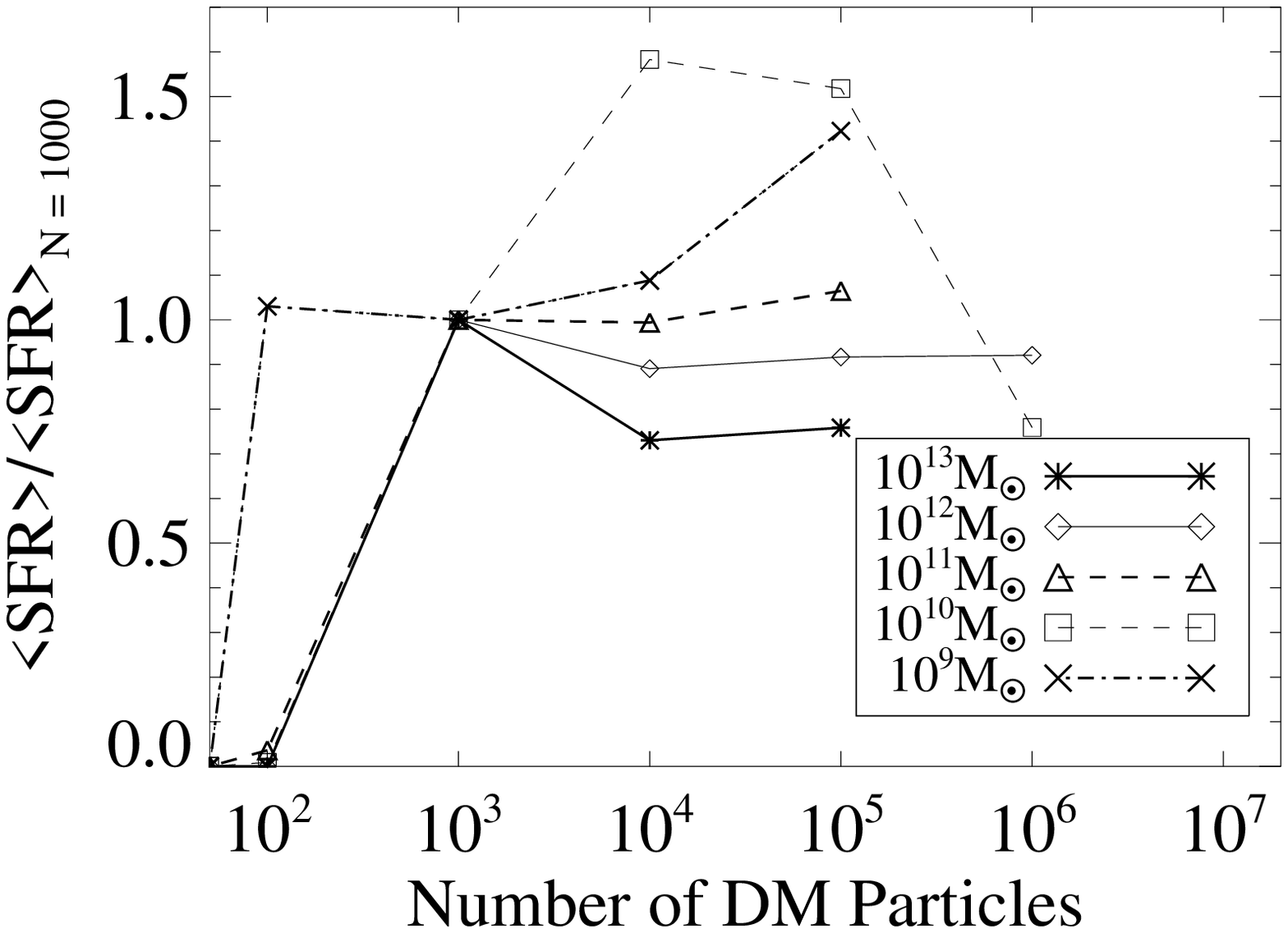}&
\\
\includegraphics[width = .45\textwidth]{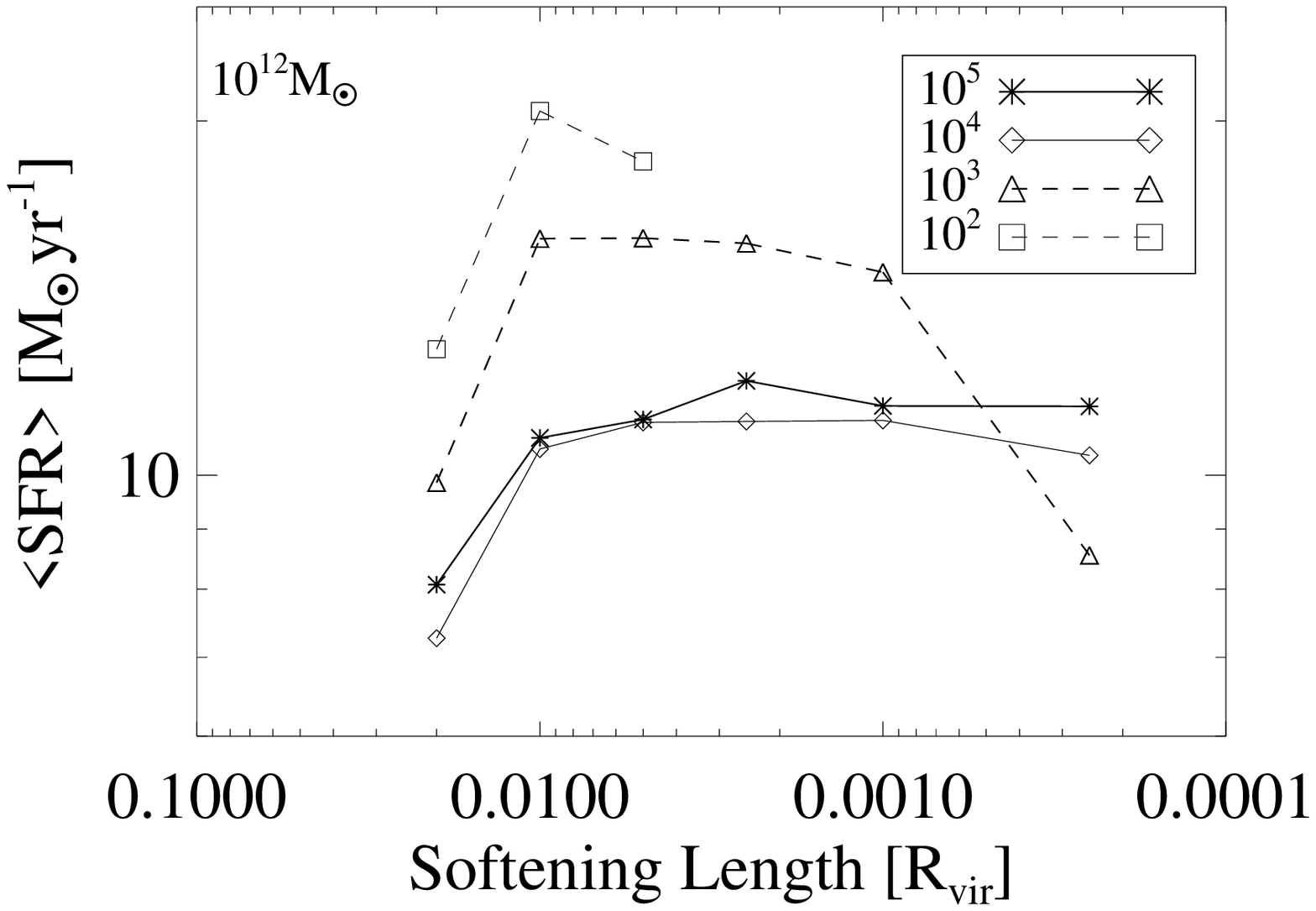}
\includegraphics[width = .45\textwidth]{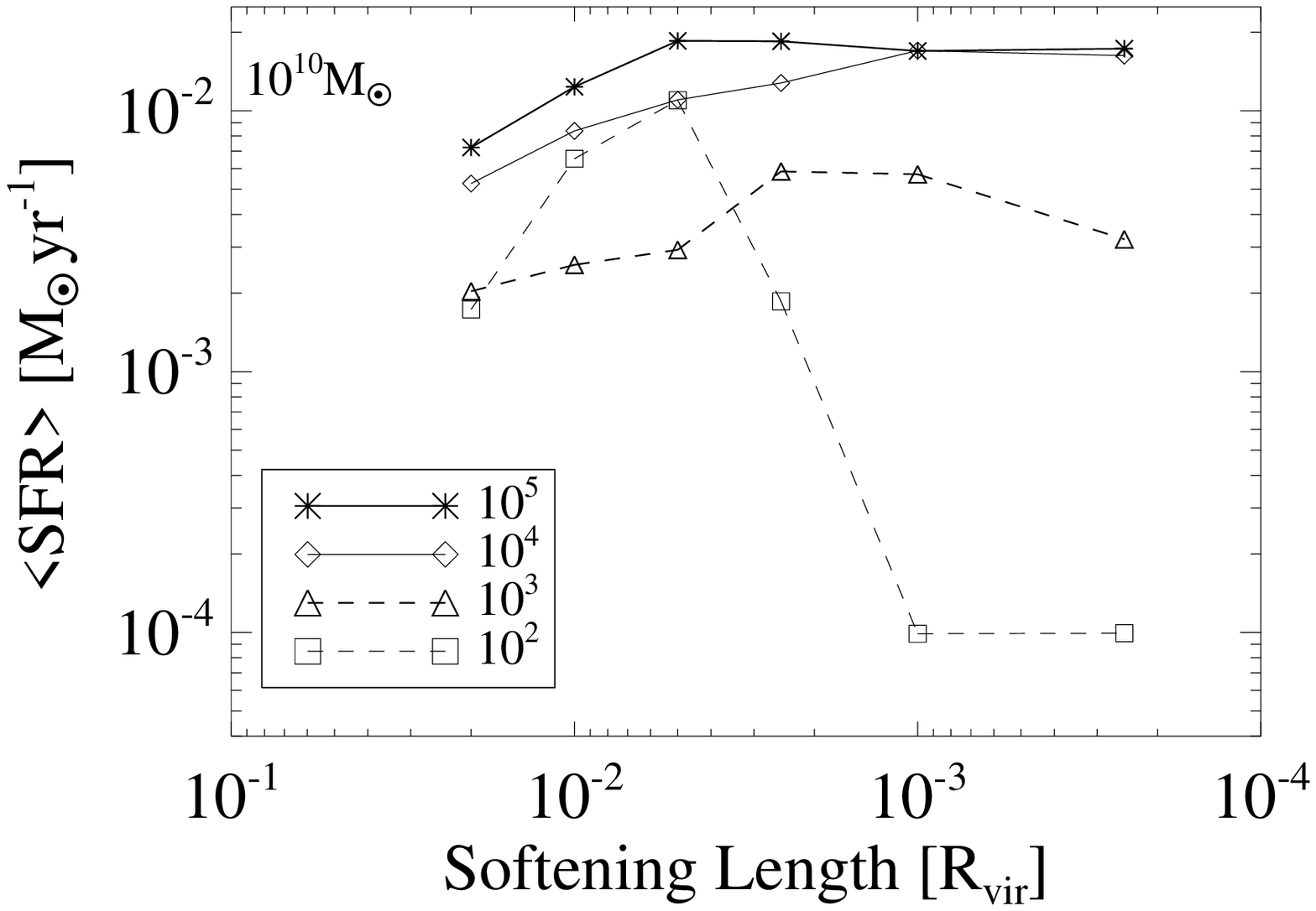}&
\end{array}$
\end{center}
\caption{Average global SFR as a function of particle number and force resolution over 3 Gyrs.   
The top panel shows the average global SFR for galaxies over a range of mass resolutions, normalized by the average global SFR of the $10^3$ DM particle simulations for each mass.
This normalization was chosen to highlight the effect of the changing mass resolution, particularly in $10^{10} M_\odot$ galaxy. 
Each curve represents a different mass galaxy and the x-axis delineates the number of DM particles used.  
For these galaxies, the force resolution is kept at $10^{-3} R_{vir}$.   
The lower two panels show the effects of both force resolution and the number of DM particles for the $10^{12} M_\odot$ (left) and $10^{10} M_\odot$ (right) galaxies.
In these panels, each line represents a different number of particles.}
\label{fig:sf}
\end{figure}

Trends in the global SFR as a function of mass and force resolution can set lower limits on the appropriate resolution for simulations of different mass galaxies.
The top panel of Figure~\ref{fig:sf} shows the average SFR over 3 Gyr for different masses of galaxies over our range of mass resolution.
For galaxies with masses of $10^{12} M_{\odot}$ or more, SFRs in simulations do not converge until $10^4$ DM particles are used; galaxies with these masses and $10^2$ DM particles or fewer do not form stars at all.
For $10^{11} M_{\odot}$ galaxies and smaller, low mass resolution simulations are still able to form stars but the SFRs differ greatly from those of simulations with high mass resolution.
These low mass, low resolution galaxies form stars in our simulations because the absolute mass resolution is still relatively high, allowing the gas to more easily reach sufficient density for our SF criteria. 
For example, during the initial collapse about 20\% of the gas particles in the $10^{10} M_{\odot}$, 100 DM particle simulation are sufficiently dense to form stars.
In comparison, none of the gas in the $10^{12} M_{\odot}$, 100 DM particle simulation is capable of SF.
In all cases, the SFR increases dramatically once 1000 DM particles are used as more of the gas particles in all the simulations are able to reach the density threshold.
The $10^{10} M_{\odot}$ galaxy is unique in showing a comparatively large increase in the SFR from $1000$ to $10^4$ DM particles. 
This increase in SFR corresponds to an increase in disk strength and will be discussed in more detail in \S 3.3.
As the number of initial gas particles is further increased to $10^6$, the SFR in this galaxy again declines. 
We will show that this can be linked to an increase in the effectiveness of stellar feedback (\S 3.2).

The bottom panels of Figure~\ref{fig:sf} show the change in the average SFR over 3 Gyr for the $10^{10} M_{\odot}$ and $10^{12} M_{\odot}$ galaxies with different numbers of particles and force resolutions.
For the simulations with the highest mass resolution, values of $\epsilon \leq 10^{-2} R_{vir}$ result in very similar SFRs.
For simulations with 1000 DM particles or fewer, however, the softening length begins to affect the SFR.
For these simulations, softening lengths at both extremes result in smaller SFRs.
Large values of $\epsilon$ reduce the SF by spreading local density enhancements over larger volumes.
While at late times this is not sufficient to limit the amount of SF, during the initial collapse it prevents the formation of a high density region in the center of the galaxy.
Because of this, about 20\% fewer gas particles are able to reach SF densities in the $\epsilon = 2 \times 10^{-2} R_{vir}$ than in the $\epsilon = 2.5 \times 10^{-3} R_{vir}$ simulation for both masses of galaxies.
Small values of $\epsilon$, in contrast, are known to result in greater two-body heating \citep{Steinmetz97}.
In our low mass resolution simulations (1000 DM particle or less) this has the effect of heating the gas and lowering the density such that fewer particles are capable of forming stars.
As the softening lengths decrease from $10^{-2}$ to $2.5\times 10^{-4} R_{vir}$, the fraction of gas that has sufficiently low temperature and high density to form stars after 3 Gyrs decreases from 17\% to 0\% in the $10^{12} M_{\odot}$ galaxy and from 11\% to 2\% in the $10^{10} M_{\odot}$ galaxy.
The global SFR of our simulated galaxies begins to converge for simulations with more than $10^4$ DM particles and for softening lengths smaller than $2.5 \times 10^{-3} R_{vir}$.  

\begin{figure}
\centering
\includegraphics[width = 0.95\textwidth]{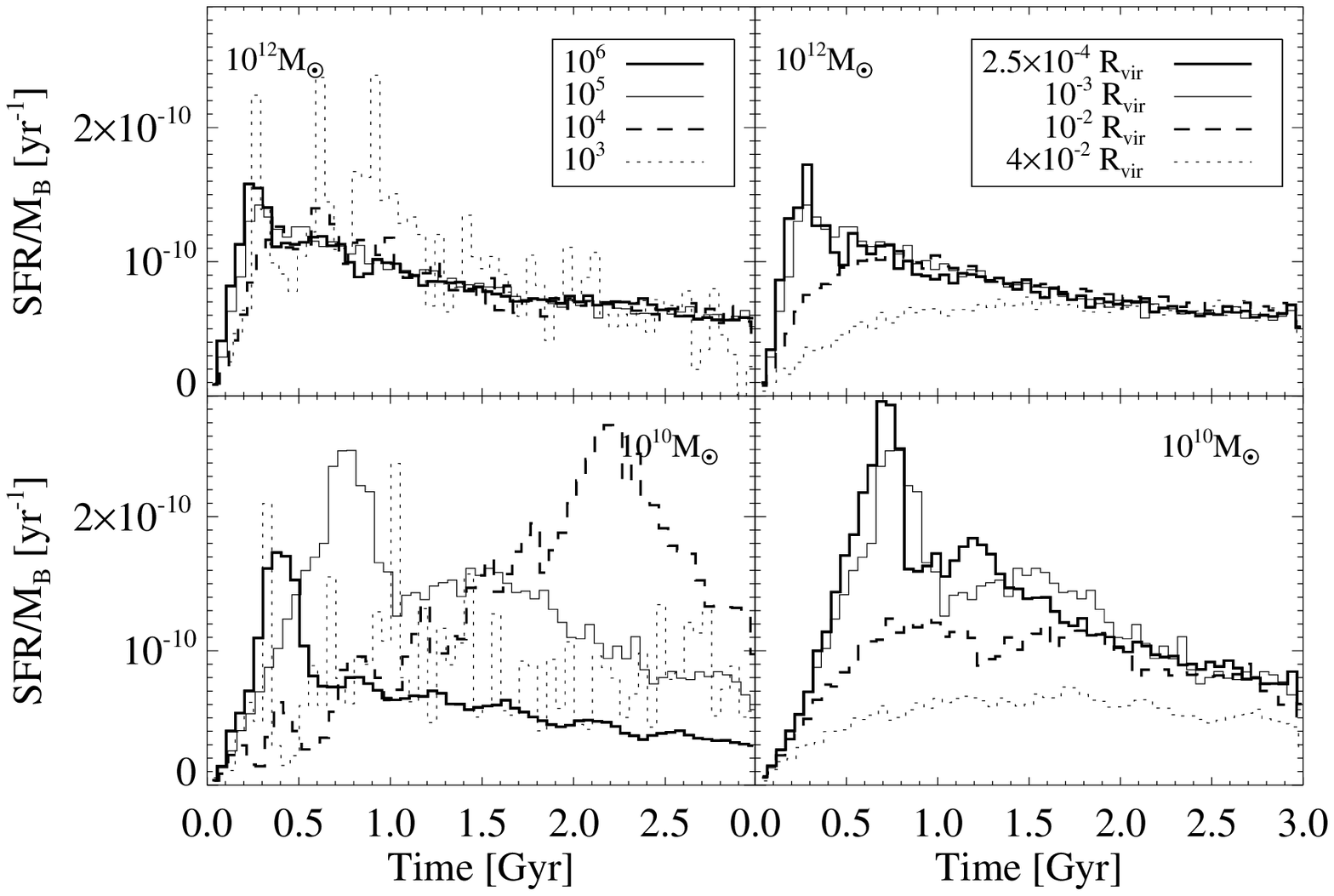}
\caption{SFHs for $10^{12} M_{\odot}$ and $10^{10} M_{\odot}$ galaxies for different force and mass resolutions.  
The SFHs are calculated using bins of $5 \times 10^7$ years and are normalized by the initial baryonic mass ($M_B$). 
 The two left panels show the effect of the number of DM particles on the SFR for a $10^{12} M_\odot$ (top) and $10^{10} M_\odot$ (bottom) galaxy with force resolution of $10^{-3} R_{vir}$.  
The two right-hand panels the show the effect of varying force resolution for a $10^{12} M_\odot$ (top) and $10^{10} M_\odot$ (bottom) galaxy with $10^5$ DM particles.}
\label{fig:sfrhist}
\end{figure}

Star formation histories (SFHs) show the changes in the global SFR as the galaxy evolves and examining them gives us an idea of the global SFR on small time scales.  
These may be observationally determined through stellar populations synthesis modeling \citep{Bruzal03} and are important for establishing the history of a galaxy and relating the SF to other process such as mergers, gas infall, disk instabilities, and stellar feedback.
In our analysis, we calculate the global SFR for every $5 \times 10^7$ years and normalize by the initial baryonic mass.
Figure~\ref{fig:sfrhist} shows both the effect of mass and force resolution on SFHs for the $10^{12} M_{\odot}$ and the $10^{10} M_{\odot}$ galaxies.  For the $10^{12} M_{\odot}$ galaxy, the SFH converges for simulations with $10^4$ DM particles or more. 
As seen in the cosmological simulations of \citet{Naab07}, in lower mass resolution simulations the initial burst of SF is delayed.
SFHs of both galaxies converge at a softening length of approximately $\epsilon = 10^{-3} R_{vir}$.
As was seen with the global SF, larger values of $\epsilon$ reduce the SFR, particularly affecting the initial burst of SF because the sudden increase in density at the core of the collapsing cloud is effectively spread over a larger volume and shallower potential.

The $10^{10} M_{\odot}$ galaxy is extremely sensitive to changes in mass resolution. 
The SFH of the $10^3$ DM particle simulation is dramatically different from all of the others and consists of a series of small bursts of SF.
 The $10^4$ and $10^5$ DM particle simulations both show a sustained burst of SF of approximately the same strength but offset in time by approximately 1.5 Gyrs.
Increasing the resolution from $10^5$ to $10^6$ DM particles decreases the strength of the burst as stellar feedback becomes more effective and the SFH qualitatively approaches the behavior of the $10^{12} M_{\odot}$ galaxy.
These sudden changes in the SFH leads to the differences in global SFRs seen in Figure~\ref{fig:sf}.
The episodic SF of the 1000 DM particle simulation may be because this galaxy does not form a disk when simulated at low resolution, as will be discussed in \S 3.3.
Instead, at low resolution this galaxy is supported in an elliptical shape by stellar feedback (\S 3.2), and the resulting SF shows a ``breathing'' pattern similar to that seen in \citet{Stinson07}.

In simulations with low mass resolution, there is a point where the large particle mass exceeds the Jeans masses of the gas particles capable of SF.
In the case of the $10^{10} M_{\odot}$ galaxy, this happens for about $10^3$ particles or fewer, and in the $10^{12} M_{\odot}$ galaxy it happens for about $10^5$ particles and fewer.
While the shift from bursty to sustained SF in the former case could be related to resolving the Jeans masses, it would have to be specific to this mass of galaxy as there is no obvious difference in the SFH for the latter case.
To address the concern that as the mass resolution increased, the higher densities of the particles and lower temperatures would result in Jeans masses for our particles that were too small to be resolved, we further examined the Jeans mass of each particle.
We found that in both the  $10^{10} M_{\odot}$, $10^5$ DM particle galaxy simulation and in the $10^6$ DM particle one, there was no difficulty in resolving the Jeans mass. 
For these two simulations, the lack of low temperature cooling kept the Jeans mass relatively high implying that the difference in SFHs could not be an effect of changing Jeans masses. 

\subsection{Stellar Feedback}
\begin{figure}
\centering
\includegraphics[width = .95\textwidth]{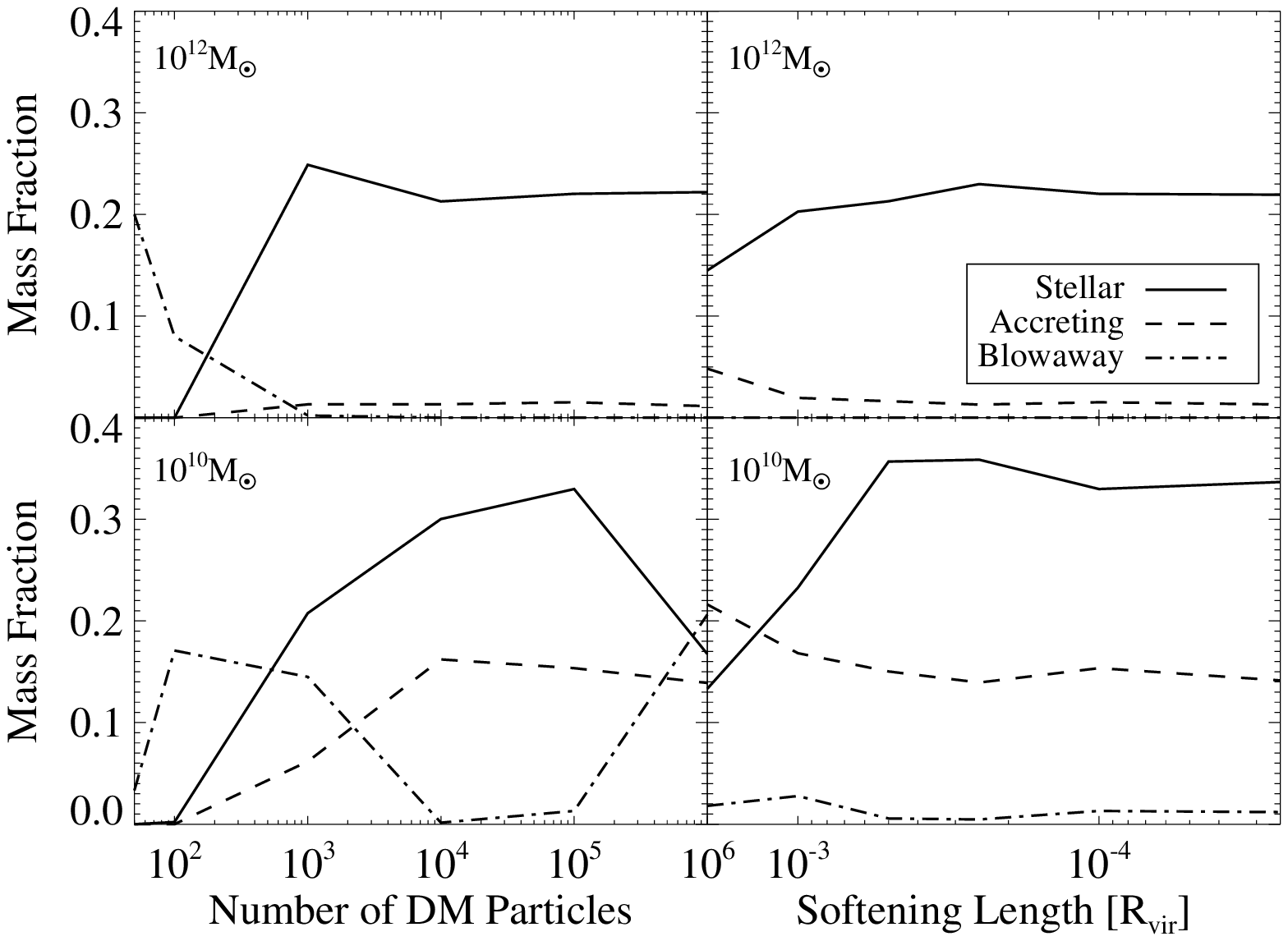}
\caption{Fate of the gas in  $10^{12}$ and $10^{10} M_{\odot}$ galaxies for different force and mass number resolutions, with data collected at 2, 2.5 and 3 Gyrs and averaged together. 
The left panels are for a $10^{12} M_\odot$ (top) and $10^{10} M_\odot$ (bottom) galaxy with force resolution of $10^{-3} R_{vir}$ and show the effect of different numbers of particles.  
The two right-hand panels are also for a $10^{12} M_\odot$ (top) and $10^{10} M_\odot$ (bottom) galaxy but with $10^5$ DM particles and varying force resolution.  
The categorizations for the gas are adopted from \citet{MacLow99}.
The accreting gas is designated as cold, dense gas moving toward the center and the stellar fraction is the relative amount of gas that formed into stars.  
``Blowaway'' denotes the fraction of baryonic mass in the form of gas permanently lost to the galaxy.} 
\label{fig:census}
\end{figure}

\begin{figure}
\centering
\includegraphics[width = .95\textwidth]{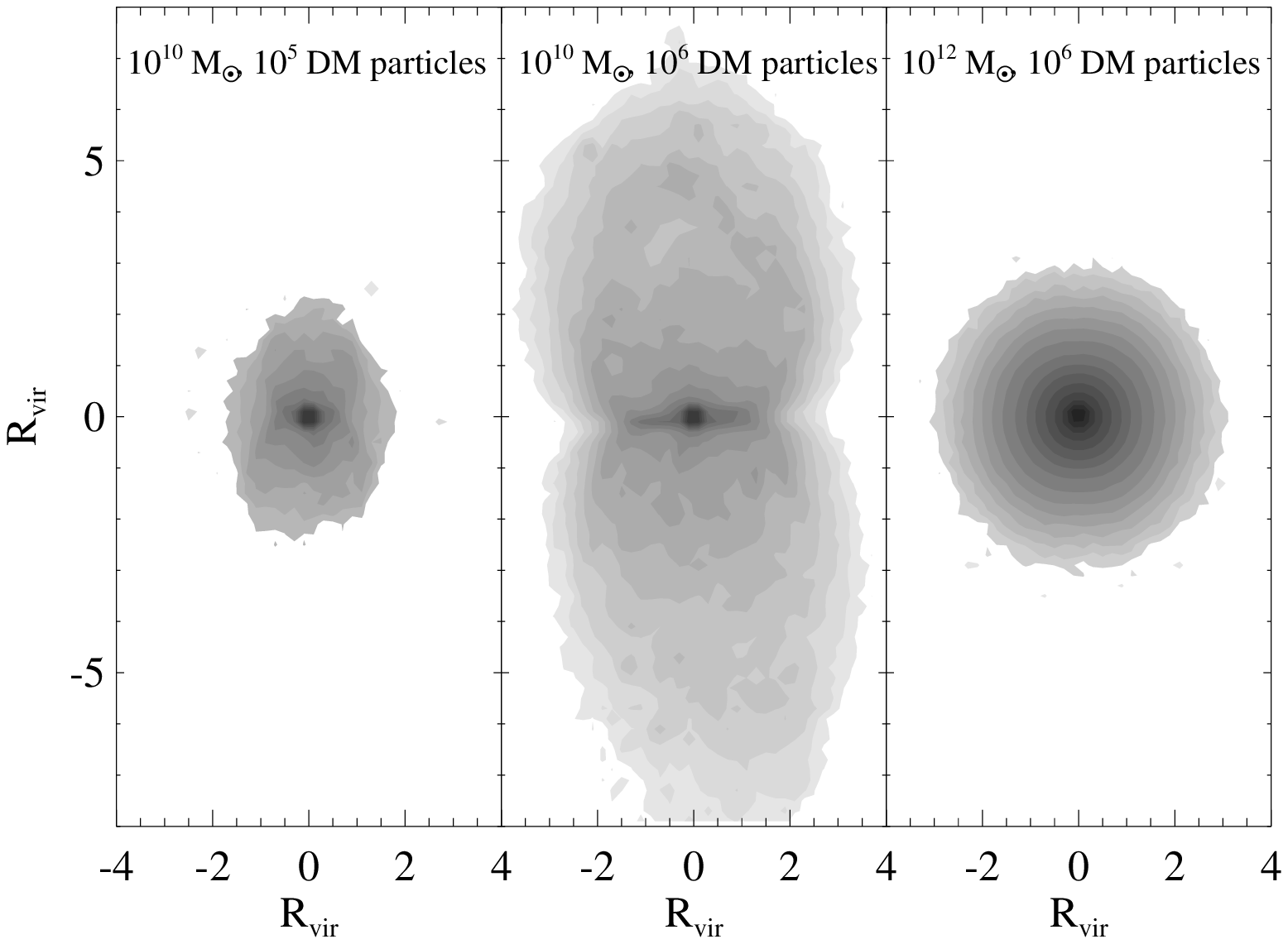}
\caption{Distribution of gas in the  $10^{10} M_\odot$ galaxy with $10^5$ DM particles (left), 
the  $10^{10} M_\odot$ galaxy with $10^6$ DM particles (middle), and the  $12^{10} M_\odot$ galaxy with $10^6$ DM particles (right) after 3 Gyrs.  
The grayscale corresponds to surface density of gas for pixel on a logarithmic scale.  
Because of different virial radii, the same color in the $10^{12} M_\odot$ galaxy represents a factor of 22 times more in total mass per pixel.  
Both simulations of the $10^{10} M_\odot$ galaxy show escaping plumes of gas caused by feedback but the $10^6$ DM particle simulation undergoes much greater mass loss.}
\label{fig:plume}
\end{figure}

The global SFR affects the structure of a galaxy through stellar feedback as energy from supernovae is injected into the surrounding media, heating the gas and, in some cases, expelling it from the galaxy.
Not only must the SFR converge but the effects of SF through feedback must also converge.
To determine the effect of feedback on gas in the galaxies, the gas was categorized as in \citet{MacLow99} according to its fate.
Figure~\ref{fig:census} shows the fractions of baryonic mass in the form of stars, accreting gas 
(cool, dense gas moving toward the center of the galaxy), 
and ``blowaway'' gas (unbound gas).
The bulk of the remaining gas resides in the disk of the galaxy.
In order to minimize any time dependency, the data was gathered after  2, 2.5 and 3 Gyrs and the fractions were averaged together.
For both masses of galaxies, softening lengths of $5 \times 10^{-3} R_{vir}$ are sufficient for convergence, representing the relative unimportance of spacial resolution on the efficiency of feedback above a certain threshold.
Mass resolution, on the other hand, has a strong effect.
As the number of particles increases, 
plumes of gas are more able to escape the disk.
The fractions of gas converge for the 1000 DM particle simulation of the $10^{12} M_{\odot}$ galaxy.

For the $10^{10}$ M$_\odot$ galaxy, however, the mean SFR does not converge even with $10^6$ DM particles. 
For this galaxy, the effect of feedback increases, causing much more gas to be expelled from the galaxy and less to be accreted.  
Figure~\ref{fig:plume} shows the large plume of gas driven from the $10^6$ DM particle galaxy.
Compared to the  $10^{12} M_{\odot}$ galaxy simulated with $10^6$ DM particles, this plume is a much larger fraction of the total mass of the galaxy.
As is expected from the comparatively shallow potential, gas is much more easily expelled from this galaxy.
This shallow potential, in combination with the changing morphology of the disk (\S 3.3), make feedback in the $10^{10} M_{\odot}$ galaxy particularly sensitive to resolution.
As the number of DM particles increases from $10^5$ to $10^6$, the mass of gas expelled sharply increases.
However, the $10^6$ DM particle simulation has less total SF than the $10^5$ DM particle simulation at almost all times (Figure ~\ref{fig:sfrhist}), indicating that the large amounts of feedback in the former are not caused by greater total amounts of SF.
Indeed, feedback appears to be suppressing SF in this galaxy.
Instead, we suggest that the different amounts of expelled gas are because of changes to the gas disk.
Increased mass resolution results in thinner disks, as will be shown in the following section. 
Additionally, using more particles results in smaller average smoothing lengths causing decreased amounts of numerical viscosity in high mass resolution simulations.
The exact effects of lowering the numerical viscosity in simulations are poorly understood, making it difficult to disentangle these two potential causes.
Both thinner disks and lower numerical viscosity, however, would be expected to aid the expulsion gas by reducing the force necessary for the winds to push out of the disk.

\subsection{Disk Structure}
While the global SF may be sufficient information to analyze large-scale simulations of structure formation, when studying individual galaxies the spatial distribution of SF is important.  
The distribution of stars and overall morphology of galaxies is observationally used to characterize galactic type, indicates the merger history of the galaxy \citep{couch98}, and is used to establish trends such as the Tully-Fisher relationship.
We examine the effects of mass and force resolution on the location of SF by looking at the properties of the stellar disks.
In particular, we relate the mass and force resolution to the resulting rotation curves, scale heights, and ellipticities of our galaxies.

\begin{figure}
\centering
\includegraphics[width = 0.95\textwidth]{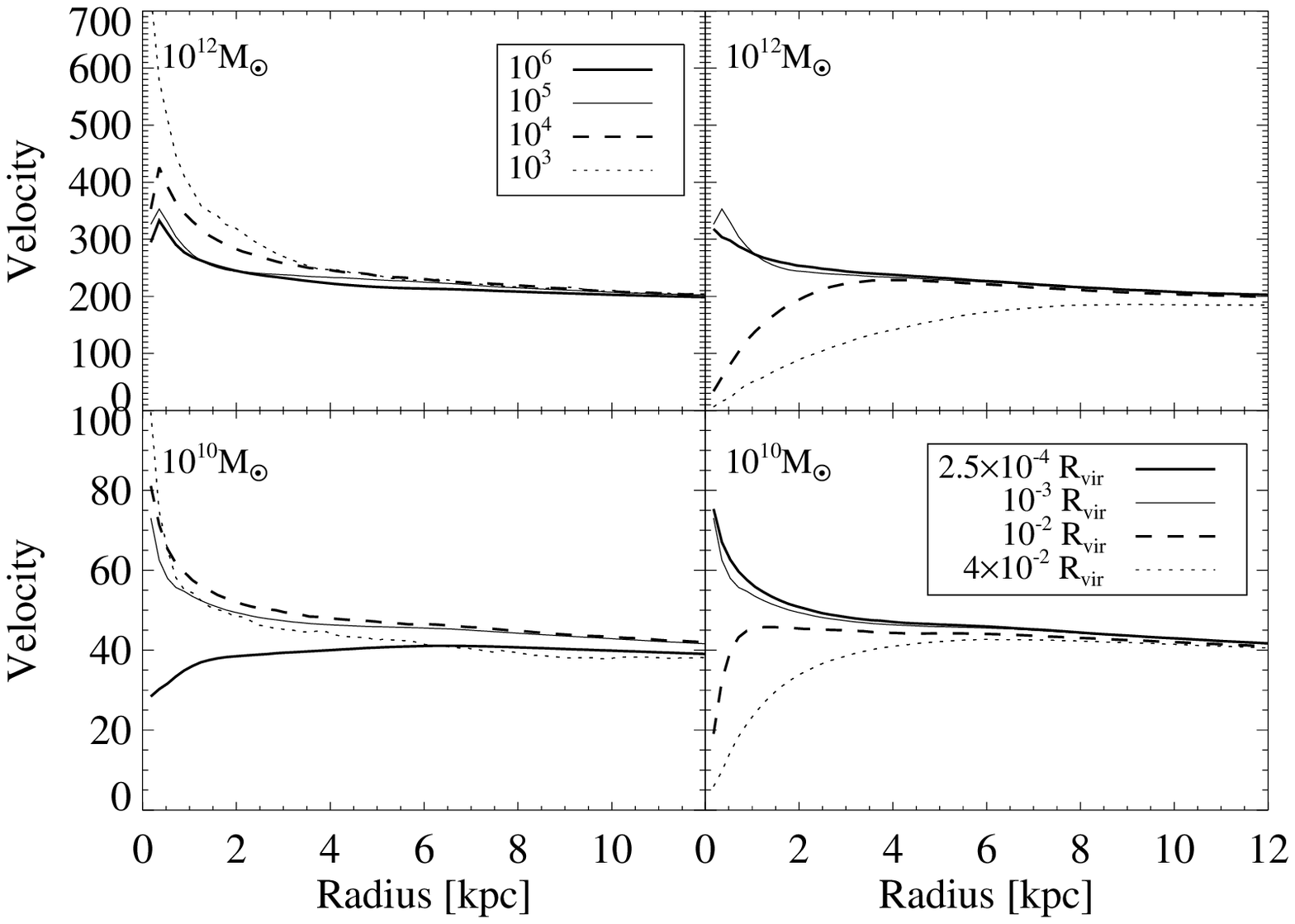}
\caption{Rotation curves of the $10^{12}$ and $10^{10} M_\odot$ galaxies for different force and mass resolutions after 3 Gyrs.  
The two left panels show the effect of the number of particles on the profiles for a $10^{12} M_\odot$ (top) and $10^{10} M_\odot$ (bottom) galaxy with $\epsilon = 10^{-3}R_{vir}$.  
The two right-hand panels show the same $10^{12} M_\odot$ (top) and $10^{10} M_\odot$ (bottom) galaxy but simulated with $10^5$ DM particles and a range of force resolutions.
}
\label{fig:profile}
\end{figure}

\begin{figure}
\centering
\includegraphics[width = 0.95\textwidth]{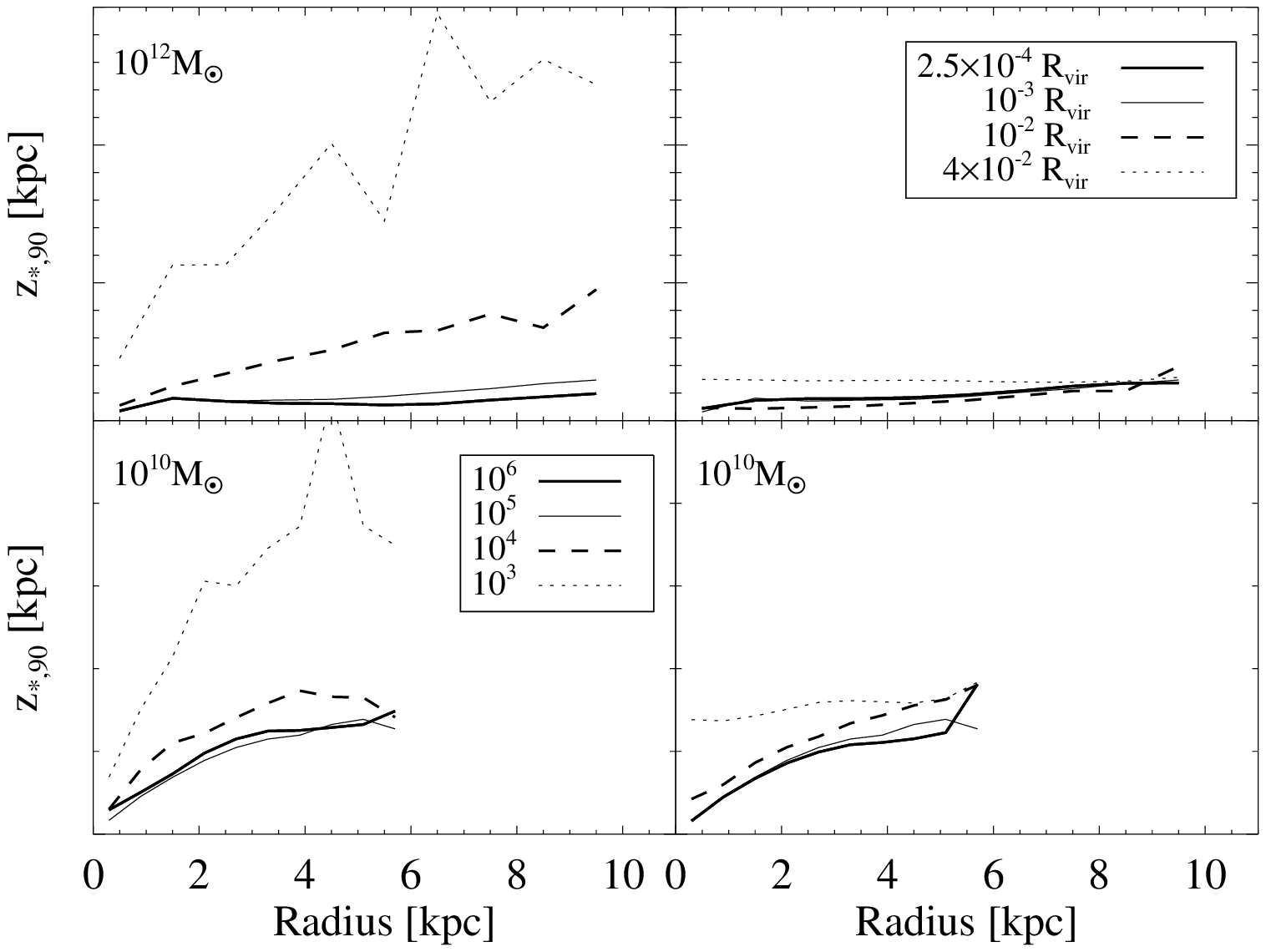}
\caption{Effect of different mass and force resolutions on the scale height of the stellar disk as a function of radius for the $10^{12}$ and $10^{10} M_\odot$ galaxies.  
The stellar scale height is defined as the height at which $90\%$ of the stellar matter is enclosed and is plotted as a function of the radius for different resolutions and masses.  
The two left panels show the scale heights after 3 Gyrs for a $10^{12} M_\odot$ (top) and $10^{10} M_\odot$ (bottom) galaxy with force resolution of $\epsilon = 10^{-3} R_{vir}$ simulated with different numbers of DM particles. 
The two right-hand panels show the effect of varying the force resolution for a $10^{12} M_\odot$ (top) and $10^{10} M_\odot$ (bottom) galaxy simulated with $10^5$ DM particles.}
\label{fig:sh}
\end{figure}

Figure 5 shows the effect of resolution on the rotation curve. 
The circular velocity, $v_c$, is defined such that $v_c = \sqrt{GM/r}$.  
Almost of our galaxies have a high central mass concentration as represented by the steep increase in $v_c$ at small radii.  
The concentrations are unaffected by increasing mass resolution.  
This contrasts with the findings of \citet{Governato07}, in which an increase in resolution similar to the increase from $10^5$ to $10^6$ particles in our $10^{12} M_\odot$ simulation resulted in a significant decrease in the inner circular velocity.
The difference likely results from the isolated environment of our simulations during which gas collapses into pre-existing DM halos and within which there are no outside influences such as mergers or tidal torques to affect the galaxy's angular momentum.  

One simulation that does show a significant decrease in the inner circular velocity is the $10^{10}$ M$_\odot$ galaxy simulated with DM $10^6$ particles.
In this galaxy, the simulation with $10^6$ DM particles is strikingly less centrally concentrated than those of lower resolution.
The low concentration is the result of the highly effective stellar feedback in this particular galaxy, as seen in the large plume of escaping gas in Figure~\ref{fig:plume}.
Similarly to the galaxies in \citet{Mashchenko2008} and \citet{Governato10}, the large amounts of stellar feedback appear to have reduced and flattened the central dark matter concentration.
Within the central 30 to 300 pc of the $10^{10} M_\odot$ galaxy, the logarithmic slope of the dark halo decreases from $-1.01$ in the $10^5$ DM particle simulation to $-0.51$ in the $10^6$ DM particle one.
A second key effect of the strong supernova-generated outflows, which is also seen in the high SF threshold cosmological simulations presented in \citet{Governato10}, is the extremely low SF efficiency at the center of the galaxies that prevents the formation of a stellar cusp.
Both the cosmological simulations of dwarf galaxies by \citet{Mashchenko2008} ($2 \times 10^9 M_{\odot}$ at z = 5) and \citet{Governato10} ($3.5 \times 10^{10} M_{\odot}$ at z = 0) included gas cooling through metal lines and a high density threshold for star formation ($\eta_{min} = 10$ amu cm$^{-3}$ and $\eta_{min} = 100$ amu cm$^{-3}$, respectively), in addition to exceptionally high mass resolution.
Despite the fact that our simulations allowed SF in lower density gas and not did not include a prescription for cooling below $10^4 K$, this galaxy demonstrates how feedback can affect the DM as well as the baryonic distribution of matter in isolated simulations.

In general, force resolution affects the rotation curves more than mass resolution.  
While low force resolution simulations result in rotation curves more similar to observed galaxies, this is a numerical artifact.
The large softening lengths simply cause the gas properties to be averaged over larger volumes and prevent a concentrated central density from forming.

As shown in Figure~\ref{fig:sh}, both mass and force resolution affect the scale height of the stellar disk. 
A decrease in the mass resolution causes the scale heights of the $10^{12} M_{\odot}$ and $10^{10} M_{\odot}$ galaxies to be larger at most radii, indicating puffier disks.
As the number of particles increases, two-body heating decreases, and thinner disks are able to form.
Varying the softening length has a negligible effect on the scale height of the disk of the $10^{12} M_{\odot}$ galaxy.
For the  $10^{10} M_{\odot}$ galaxy, on the other hand, lower force resolution has a similar effect to lower mass resolution: it increases the scale height.
The puffier disks at low force resolution are produced as the shallower potential allows the feedback to become more effective.
This appears to be another situation in which the $10^{10} M_{\odot}$ galaxy is very sensitive to numerical parameters and is connected to the wide range of feedback efficiencies seen in this mass of galaxy.

\begin{figure}[ht]
\begin{center}$
\begin{array}{cc}
\includegraphics[width = .45\textwidth]{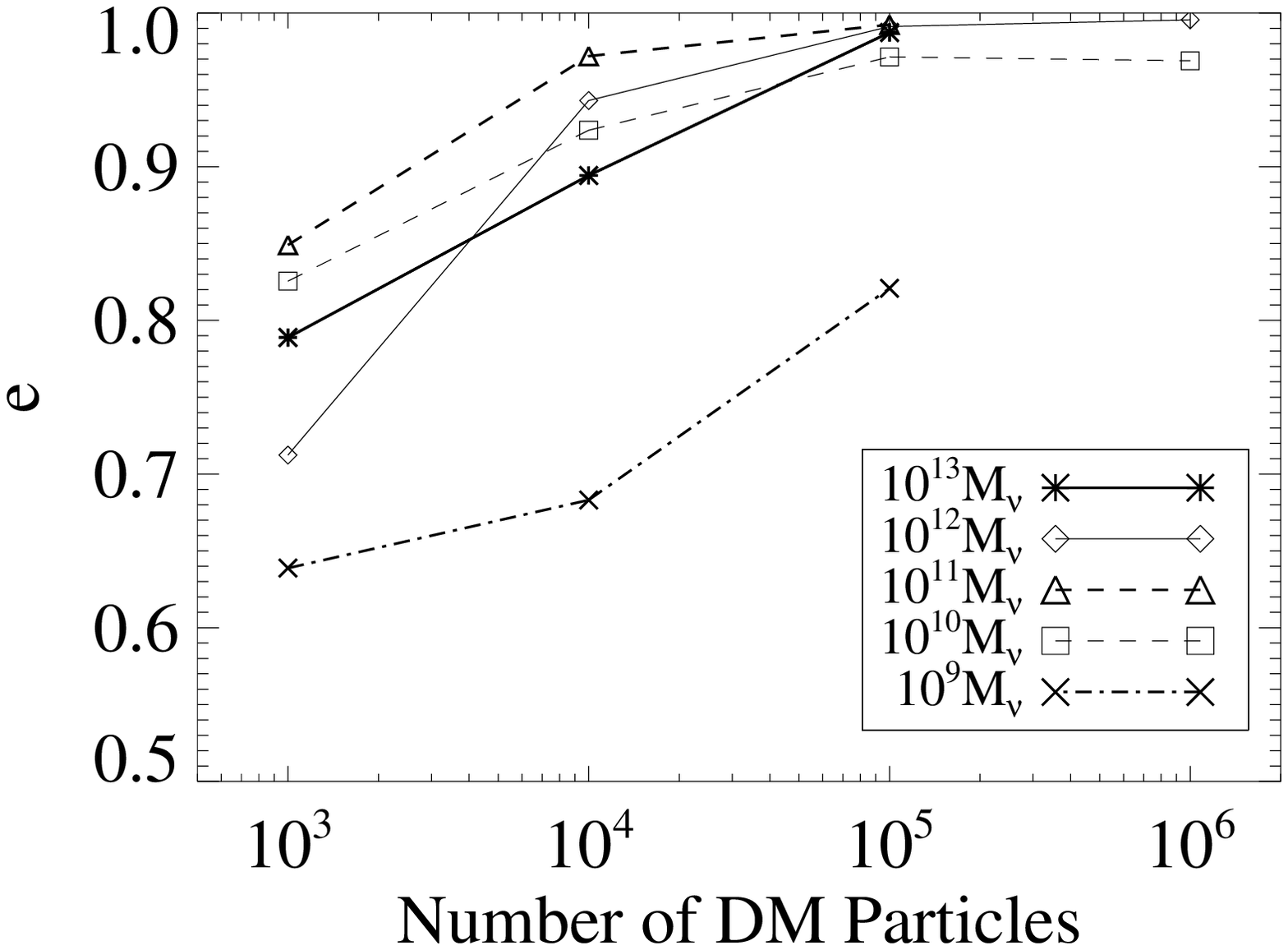}&
\\
\includegraphics[width = .45\textwidth]{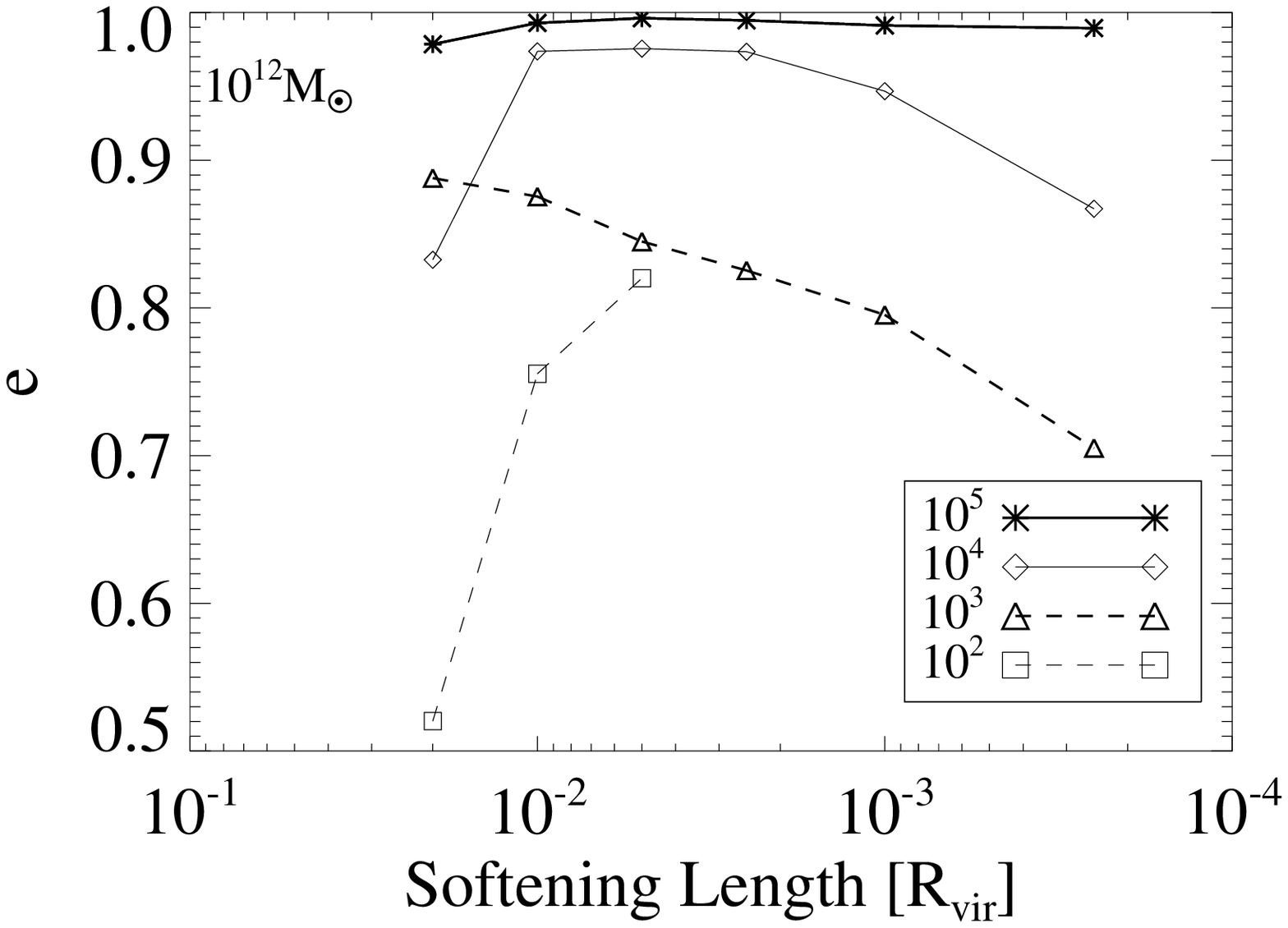}
\includegraphics[width = .45\textwidth]{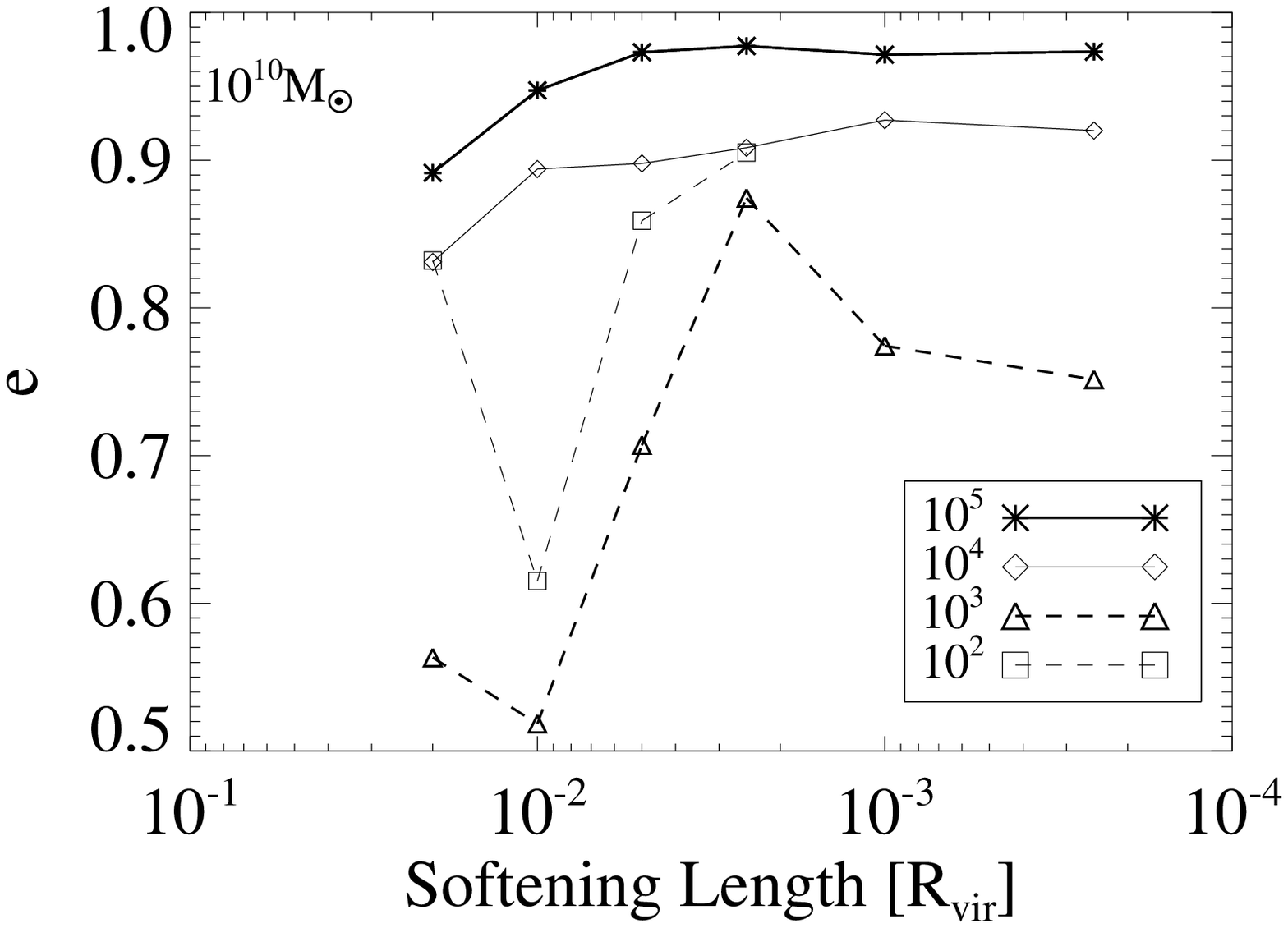}&
\end{array}$
\end{center}
\caption{Ellipticity ($e$,  defined for the stellar matter as in \S 3.3) of the galaxies viewed edge on after 3 Gyrs.  
In the upper panel, the ellipticity is plotted against the number of DM particles in the simulation for each mass of galaxy.  
The lower panels show the effect on the ellipticity of changing the spacial resolution for a $10^{12} M_{\odot}$ (left) and a $10^{10} M_{\odot}$ (right) galaxy with $10^5$ DM particles.} 
\label{fig:ellip}
\end{figure}

In order to combine the radial stellar disk profile and the scale height of the disk into a single parameter representing disk strength, we calculate the ellipticity of the stellar distribution as viewed edge on.
To do this, we chose the major axes to be $r_{*,90}$, the radius at which $90\%$ of the stellar mass is enclosed, and the minor axes to be $z_{*,90}$, the height at which $90\%$ of the stellar mass is enclosed.
Using $r_{*,90}$ and $z_{*,90}$, we defined the ellipticity to be $e = \sqrt{\frac{r_{*,90}^2 - z_{*,90}^2}{r_{*,90}^2}}$, where $e = 1$ is a disk and $e = 0$ is a sphere.
Using the ellipticity of the galaxy, measured in this way, as a proxy for the strength of the disk is an ad hoc method and was chosen simply as a straight forward way of quantifying the galaxy's appearance.
The upper panel of Figure~\ref{fig:ellip} shows the ellipticity of galaxies of different masses and mass resolutions.  
Across all resolutions, the $10^{9} M_{\odot}$ galaxy has a more spherical shape.  
The $10^{10} M_{\odot}$ galaxy is the lowest mass galaxy to form a disk, the formation of which is strongly dependent on the number of particles in the simulation.  
In general, the strength of the disk (i.e.  greater ellipticity) increases sharply with mass resolution.
For the $10^{10} M_{\odot}$ galaxy, the ellipticity also roughly increases with force resolution (Figure~\ref{fig:ellip}) as the more concentrated potential causes the disk scale height to decrease.
The ellipticities of the $10^{12} M_{\odot}$ galaxy simulated with $10^3$ and $10^4$ DM particles, however, turn over at the highest force resolution (Figure~\ref{fig:ellip}).  
This is the result of the decreasing disk scale lengths at high force resolutions while the scale heights remains largely unaffected by changes to the force resolution.
The disk parameters, including rotational curves, scale height, and ellipticity, converge for the $10^{10} M_{\odot}$ and $10^{12} M_{\odot}$ galaxies for softening lengths between $10^{-2}$ and $10^{-3} R_{vir}$.
Generally, disk parameters converge in simulations of $10^5$ DM particles and higher.

\section{Discussion} \label{discussionsec}
In our simulations, we demonstrate a close relationship between resolution,
stellar feedback, spatial distribution of star formation, and the amount of star formation.
Cooling in combination with angular momentum produces disks \citep{OstrikerANDBinney89}.
Angular momentum is artificially reduced in simulations with few particles,
shrinking the size of disks formed if not preventing disks from forming entirely
\citep{Governato04}.
The higher resolution models create the thinnest (i.e. highest-ellipticity) disks, and form the most stars due to the presence of the highest density gas.
Increasing the resolution results in sustained, constant SF with a feedback-driven plume of gas.
In high-mass galaxies, this occurs in simulations with $10^3$ particles where the SF and the effects of feedback begin to converge.
In low-mass, lower-resolution galaxies with thick disks (i.e. low-ellipticity), SF is episodic as feedback shuts off SF for short periods of time. 
For these galaxies, it is most likely a combination of the relatively large disk scale height, the more spherical structure, and the greater artificial viscosity that prevent the heated gas from puncturing through the galaxy in the form of a galactic wind \citep{MacLow88, MacLow89}.  
Instead, stellar feedback causes a temporary halt to SF in the galaxy \citep{Stinson07}.

The $10^{10} M_{\odot}$ galaxy is at the transition between these two behaviors and its shallow potential makes it particularly sensitive to stellar feedback.
Increasing the mass resolution from $1000$ to $10^4$ DM particles results in the formation of a disk.
Further increasing the number of DM and initial gas particles from $10^5$ to $10^6$ produces increased gas loss through stellar feedback.
At our highest mass resolution, feedback is efficient enough to decrease the central concentration of the halo and to produce a slowly rising rotation curve.
Over the same range of mass resolution, the SFH goes from small episodic bursts, to one extremely large burst, to a smaller burst followed by sustained SF.
These changes appear to be linked; low resolution causes the galaxy to be less disky, making supernova feedback expand the gas spherically outwards and ending SF until the gas is able to one again cool and compact.
At the highest resolution, the galaxy contains a strong disk and outflows of gas concentrated along the poles.
While these outflows do not cause a galaxy-wide halt to SF, they do lower the concentration at the center of the galaxy.
We expect that it is the combination of smaller disk scale-height and lower artificial viscosity in the higher mass resolution simulations that promote the formation of the outflows.
The force resolution also has relatively strong effects on this galaxy, as large values of $\epsilon$ cause thicker disks.
In fact, even at our highest resolution, we were unable to establish the convergence of the SFR, SFH, or the fate of the gas.
This sensitivity extends even beyond resolution: changing the initial mass function from Kroupa \citep{Kroupa93} to Miller-Scalo \citep{Miller79} produced similar changes to the SFH.
The susceptibility of simulations of this mass of galaxy to small changes in the parameters is at least in part due to its low virial temperature; its virial temperature of approximately $2.5 \times 10^4$ K is close to $10^4$K, which is the temperature regime at which metal line cooling becomes important.
The set of simulations examined in this paper does not include cooling from metals or $\mathrm{H_2}$ and low temperature cooling is necessary to study these low-mass galaxies in greater depth.

There is some evidence that the strong sensitivity of this mass of galaxy is not unique to our simulations but may be based on a physical reality.
\citet{Strigari08} found that the Milky Way satellites share a common mass of $10^{7} M_{\odot}$ within their central 300 parsecs despite covering a $10^4 L_{\odot}$ range in luminosity.
Simulations predict that each of these satellites originated from galaxies with a total mass prior to interacting with the Milky Way of approximately $10^{9} M_{\odot}$ \citep{Strigari08}.
Despite the similar halo mass, differences in history and environment were capable of causing extremely different SFHs.
Although we are considering galaxies a factor of 10 more massive, it is difficult to match the masses, both because of the extrapolation of the total halo mass from the observed mass within 300 pc and the extrapolation from our isolated simulations to a satellite halo.
Therefore, the observed $10^{9} M_{\odot}$ galaxies could be in the same precarious position as our $10^{10} M_{\odot}$ galaxies.
If such a wide range of luminosities in the Milky Way satellites can be produced by similar initial mass galaxies undergoing different histories, it is possible that the sensitivity of our $10^{10} M_{\odot}$ reflects the sensitivity of real galaxies of similar mass.
Our studies demonstrates that slightly different amounts of feedback and ease of mass loss can completely change the structure and SFH of this mass of galaxy.  
While in this study, those differences were the result of changes to resolution, these could also be caused by differing merger and gas accretion histories.

One question when examining these results is how universal the requirements for convergence are.
In particular, will the same resolution be necessary in galaxies created from different initial conditions, galaxies using different star formation parameters, and galaxies that are evolved with different feedback recipes?
Similar mass resolution was required for convergence of the total amount of SF in cosmological runs of Milky Way mass galaxies \citep{Brooks07, Governato07}, with slightly different feedback energies and SF efficiencies,  implying that the resolution necessary for convergence is mostly independent of the initial conditions and of small changes in those two SF parameters.
Feedback regulated SF tends to reach an equilibrium with increased SF efficiency resulting in stronger feedback which then hinders the formation of more stars.
Based on the relative insensitivity to these parameters, we can assume that similar mass resolution would be necessary even with different values for these parameters.
The minimum density for SF, $\eta_{min}$, is the single parameter one would expect to be strongly linked to resolution.
As lower mass and spatial resolution results in less high density gas, increasing $\eta_{min}$ requires resolving the structure of the ISM to a greater degree \citep{Governato10}.
Simulations with high values of $\eta_{min}$, therefore, should need greater resolution for convergence.
A more complete discussion of the relationship between efficiency, $\eta_{min}$ and resolution may be found in \citet{Saitoh08}.

A more difficult issue is the robustness of our results across different feedback recipes.
Although most low-resolution SPH simulations tend to have depressed amounts of SF caused by the presence of a smooth, thick gas disk, the actual mass resolution necessary for convergence appears to be highly dependent on the feedback recipe. 
For example, in the \citet{SpringelANDHernquist03a} recipe used in GADGET individual gas particles represent a multi-phase ISM and some of the energy from feedback is transferred to selected nearby gas particles in the form of kicks to their velocities.
As seen in Figure 12 of \citet{SpringelANDHernquist03a}, this approach has the advantage of being relatively insensitive to mass resolution.
By imposing an equation of state upon the gas particles, self-regulation of SF is assumed; by allowing sub-grid multi-phases of the ISM within gas particles, different phases of the ISM need not be resolved.
The cost of these advantages, however, is ease in describing non-equilibrium feedback configurations.
For SF to be shut off or for superbubbles to be formed, a model for galactic winds must be assumed.
One approach when modeling galactic winds is to temporarily decouple the winds hydrodynamically from the galaxy \citep{SpringelANDHernquist03a}.
While this makes the simulations largely insensitive to resolution, the results are very sensitive to the model parameters.
In models in which the wind does remain coupled to the gas, decreasing resolution induces greater mass loss in the winds and, in more massive galaxies, causes the winds to have less effect on the disk structure \citep{DallaVecchia08}.
Despite the differences in feedback recipes, our results agree qualitatively in that higher mass and spatial resolution results in stronger disk structure and (aside from the $10^{10} M_{\odot}$, $10^6$ DM particle halo) higher SF.
However, while \citet{SpringelANDHernquist03a} was able to establish convergence of the SFR of the $10^{10} M_{\odot}$ halo, our simulations did not.
We believe this is because our simulations of this galaxy are in highly non-equilibrium states between star formation and supernova feedback, whereas some degree of self-regulated star formation is assumed in the \citet{SpringelANDHernquist03a} simulations. 
We find in general that for cases other than the $10^{10} M_{\odot}$ halo, we need at least a factor four higher mass resolution in our simulations for convergence of the global SFR than in \citet{SpringelANDHernquist03a,SpringelANDHernquist03b}.

The effect of resolution on global SF and disk structure has implications when studying SF in cosmological simulations. 
Cosmological simulations generally use the same mass of particles and the same spatial resolution throughout. 
This causes less massive halos to be made-up of relatively few particles, each of whose softening lengths are a larger fraction of the virial radius, and results in less accurate SF for these halos.
Several cosmological simulations using the \citet{SpringelANDHernquist03a} feedback recipe or modifications thereof have been used to study the cosmological SFH \citep{SpringelANDHernquist03b,ChoiNagamine09,Schaye09}.
In these studies, the authors balance the need to simulate a large volume at z = 0 with the importance of high mass resolution at high redshift by stitching together a series of simulations of increasing size and decreasing mass resolution at lower redshifts.
Over each redshift interval, the cosmological SFRs within simulations of different mass resolutions are checked for convergence. 
\citet{SpringelANDHernquist03b, Schaye09} noted that while low resolution results in a decrease in the SFR at high redshifts, the amount of free gas left over may result in artificially high SFRs at z=0.  

One example of a large, cosmological SPH simulation with similar SF and feedback recipes  is \citet{Keres05}.
In their primary simulation, \citet{Keres05} simulated galaxy formation within a $22.222$ h$^{-1}$ Mpc  comoving cube using the star formation and feedback recipes from \citet{Katz96}.
Stars formed probabilistically from eligible gas particles and supernova feedback energy is transferred to the surrounding gas particles in the form of heat.
This feedback recipe is similar to ours in that it took place over single ISM phase gas particles and did not assume regulated SF, but the lack of the blastwave formulation would have caused the feedback-heated gas particles to cool more quickly.
According to our tests, in the \citet{Keres05} primary cosmological  simulation (L22/128), galaxies with masses of $10^{11} M_{\odot}$ and above were simulated with a sufficient number of particles to judge the global SFR.
For less massive galaxies, however, the mass resolution was low enough to throw into question their SFR.
In their simulation, galaxies with masses close to $10^{10} M_{\odot}$ were simulated with roughly 100 particles.
Our isolated galaxies of similar mass and particle number have 100 times less SF than their high-resolution counterparts over three billion years, although the difference decreases after the initial collapse and burst of SF.
We estimate the effect of using low mass resolution on low mass galaxies when calculating the total SFR in a simulation by multiplying the equilibrium SFR in each individual mass galaxy by the number of those galaxies expected from the galactic mass function \citep{Reed07}.
When comparing the total SFR after the initial collapse (from 1.5 to 3 billion years) generated by our best resolved galaxies to that generated by galaxies with particle masses of $10^{8} M_{\odot}$ (close to L22/128 resolution), we find our well-resolved simulations to produce about 10\% more SF.
The net difference in SF is significantly smaller than the discrepancy for low mass galaxies because most stars form in $L_*$ galaxies, which were simulated at higher resolution.
It should be noted, however, that at higher redshift, low-mass galaxies make up a higher percentage of the overall mass and the fraction of SF that is poorly constrained would increase.
Furthermore, the relationship between resolution and the global SFR, particularly in the low-mass galaxies, is not a linear one and intermediate resolution simulations could actually produce greater discrepancies in total SFRs.

Because of the difficulties in comparing simulations of isolated galaxies to cosmological simulations as well as the difference between the feedback recipes used in the simulations,  this study does not serve as a direct estimate of the possible error in the SFR.
As seen in \citet{Mayer08} and \citet{Governato08}, excess angular momentum loss may occur at different resolutions in cosmological simulations versus isolated galaxies.
Without the effects of large-scale tidal torques from cosmological simulations, our isolated galaxies may not have the same angular momentum, morphology, or disk structure.
Furthermore, the faster radiation of feedback energy in \citet{Keres05} would result in decreased regulation of SF, particularly at the low mass end, and their feedback recipe may be affected differently by resolution.
This exercise does highlight, however, the importance of resolution in  cosmological simulations and illustrates how knowledge of the effects of resolution over a range of masses can be used to generate approximate limits for the analysis of large scale simulations.

\section{Conclusions}   \label{conclusionsec}
With limited computational resources, a compromise is always necessary between sampling a large number of galaxies and resolving those galaxies well.
The practical question is not at what level of resolution do all measurements of the galaxy converge, but what minimum resolution is needed to answer a specific question.
Studying the bright end of the galactic luminosity function, for example, requires very different resolution than studying the SFHs of dwarf galaxies.

Based on our isolated galaxies, we recommend resolutions for various modeling goals in cosmological SPH simulations for the S06 SF and feedback recipes.
These recommendations may also be applicable to any SPH simulation with single-phase gas particles in which supernova feedback energy is transferred to the surrounding gas particles.
As these are based on isolated galaxies, they come with a number of caveats when extrapolating to cosmological simulations.
Namely, our galaxies have purely spherical accretion, have no mergers, and have no outside tidal forces.
These limitations affect the properties of accreted gas as well as the amount of angular momentum transferred to and from the galaxy.
Our initial conditions also only sample one standard angular momentum ($\lambda$ = 0.039) and one mass profile (NFW) and assume the \citet{Kroupa93} initial mass function.
Despite these caveats, our set of simulations covers a wide range of masses, mass resolutions, and force resolution and produce both ``bursting" and sustained SF at different stages.
Together they provide a point of reference for determining the necessary resolution of galaxies.
With these limitations in mind, we make the following recommendations.

\begin{itemize}
\item \emph{For simulations similar to ours with gas particle masses less than $10^7 M_\odot$, the Jeans mass of star forming gas is resolved}
This implies that our values of $n_{min}$ and $c^*$ in  the SF recipe are suitable for our range of resolutions using our cooling recipe. 

\item \emph{Fewer than $10^4$ DM particles or initial gas particles or softening and smoothing lengths greater than $10^{-3} R_{vir}$ do not produce accurate global SFRs and supernova feedback.}
Over a range of masses, we find that the average global SFR converges for isolated galaxy simulations of $10^4$ particles and for softening lengths smaller than $10^{-3} R_{vir}.$
Gas loss and accretion also converge for similar values (except in the $10^6$ DM particle, $10^{10} M_\odot$ simulation).
While the connection between mass loss and SF is clear, it is remarkable that the mass loss convergence happens at the same values over almost the entire range of galactic masses studied, despite the differences in the galactic potentials.

\item \emph{SFHs for galaxies do not converge for simulation using less than $10^4$ DM particles and less than $10^4$ initial gas particles or for simulations with softening lengths larger than  $10^{-3} R_{vir}$. 
In galaxies near $10^{10} M_\odot$, the interaction between feedback and the shallow potential wells create spurious effects and prevents convergence at all mass resolutions.}
In general, increasing mass resolution results in more peaked bursts of SF followed by a smooth decrease in the SFR.
The $10^{10} M_{\odot}$ galaxy, however, shows a wide range of behavior over the mass resolutions tested with no clear convergence.
For this mass of galaxy, raising the number of particles causes the creation of a disk and raising it further results in a sudden increase in the amount of mass lost through supernova feedback. 
The SFHs of this particularly sensitive galaxy reflect this: at low mass resolution the lack of a coherent disk results in episodic SF; slightly better mass resolution simulations show higher peaked SFRs; the highest resolution galaxy shows less SF as stellar feedback becomes more efficient.

\item  \emph{The spatial distribution of stars is not robust for softening lengths of $ \epsilon \ge 10^{-2} R_{vir}$ or in simulations with fewer than $10^5$ particles.}
Low mass and force resolution affect the spatial distribution of the stars by making the galaxy less disky with a less steep central density profile.
Improving the mass resolution reduces the angular momentum loss and shrinking the softening length similarly reduces artificial pressure and allows disks to collapse more fully.
Disk properties in galaxies with $10^4$ particles are only starting to converge and $10^{5}$ particles are generally necessary for the results to be robust.
\end{itemize}

For galaxies much smaller than $10^{10} M_{\odot}$, low temperature cooling from metals \citep{Mashchenko2008} and $\mathrm{H_2}$ is necessary for gas to reach temperatures below the virial temperature.
This is particularly important when predicting the effect of stellar feedback in dwarf galaxies or at high redshift.
High resolution simulations in combination with low temperature cooling and $\mathrm{H_2}$ formation would better allow us to probe the star formation within dwarf galaxies, including the effects of feedback on rotation curves and episodic versus sustained star formation.
In future work, we will examine the effect of $\mathrm{H_2}$ on cooling in galaxies of different masses and present an $\mathrm{H_2}$-based SF model

\acknowledgments
We would like to thank the anonymous referee whose comments added substantially to this paper.
C.C. was partially supported through the National Science Foundation Graduate Research Fellowship Program and Graduate Teaching Fellowship in K-12 Education. T.Q. and G.S. were supported by NSF ITK grant PHY-0205413.
The Condor Software Program (Condor), on which most of these simulations were run, was developed by the Condor Team at the Computer Sciences Department of the University of Wisconsin-Madison.  All rights, title, and interest in Condor are owned by the Condor Team.  All simulations of galaxies were computed on machines funded by the Student Technology Fee of the University of Washington.


\bibliography{./respaper}







\end{document}